# Pressure-tailored lithium deposition and dissolution in lithium metal batteries

Chengcheng Fang[1,2,†,*], Bingyu Lu[1,†], Gorakh Pawar[3], Minghao Zhang[1], Diyi Cheng[1], Shuru Chen[4], Miguel Ceja[1], Jean-Marie Doux[1], Mei Cai[4], Boryann Liaw[3], Ying Shirley Meng[1*]

[1]Department of NanoEngineering, University of California San Diego, 9500 Gilman Drive, La Jolla, CA 92093, USA.

[2]Department of Chemical Engineering and Materials Science, Michigan State University, 220 Trowbridge Rd, East Lansing, MI 48824, USA.

[3]Energy and Environmental Science and Technology Directorate, Idaho National Laboratory, 1955 N. Fremont Avenue, Idaho Falls, ID 83415, USA.

[4]General Motors Research and Development Center, 30470 Harley Earl Blvd., Warren, MI 48092, USA.

[*]Correspondence to: shirleymeng@ucsd.edu (Ying Shirley Meng), cfang@msu.edu (Chengcheng Fang)

[†]These authors contributed equally

**Abstract:** A porous electrode resulting from unregulated Li growth is the major cause of the low Coulombic efficiency and potential safety hazards of rechargeable Li metal batteries. Strategies aiming to achieve large granular Li deposits have been extensively explored; yet, the ideal Li deposits, which consist of large Li particles that are seamlessly packed on the electrode and can be reversibly deposited and stripped, have never been achieved. Here, by controlling the uniaxial stack pressure during battery operation, a dense Li deposition (99.49% electrode density) with an ideal columnar structure has been achieved. Using multi-scale characterization and simulation, we elucidated the critical role of stack pressure on Li nucleation, growth and dissolution processes, and developed innovative strategies to maintain the ideal Li morphology during extended cycling. The precision manipulation of Li deposition and dissolution is a critical step to enable fast charging and low temperature operation for Li metal batteries.

**Main text:**

Lithium (Li) metal is the ultimate anode material to break the specific energy bottleneck of going beyond Li-ion batteries. However, due to its low Coulombic efficiency (CE) and safety issues caused by possible dendrite growth and inactive Li formation, practical rechargeable Li metal batteries have not yet been realized since its inception in 1976 (*1–3*). It is widely accepted that the morphology is one of the determinantal factors for CE and cycle life of Li metal batteries (*4, 5*). In order to achieve dense Li deposition close to the actual density of Li metal (0.534 g/cm$^3$), tremendous efforts have been devoted to understanding and controlling the Li deposition process by considering the electroplating as a mass-transport controlled process, which is primarily affected by factors including electrolyte properties (cation concentration, solvation structure, etc.), current density and temperature (*6, 7*). In addition, due to the highly reducing potential of Li, the (electro)chemically formed solid electrolyte interphase (SEI) between the Li metal and liquid electrolyte makes the electroplating a kinetically slow solid diffusion process.

Thus, the Li deposition and dissolution are further affected by the SEI properties. Accordingly, strategies aiming to improve the Li metal anode performance have been extensively designed to favor at least one of the four governing factors in the past decades: e.g. 1) engineering the electrolyte towards large granular Li particle deposition and stable SEI (*8–10*) 2) utilizing 3D current collectors to increase surface areas thus to reduce local current density (*11*, *12*); 3) creating artificial SEI to facilitate Li ion transport and prevent parasitic reactions (*13*); and 4) applying elevated temperature to enhance the mass transfer for enlarged Li particles growth (*14*). However, the multidimensional commercial requirements of Li metal batteries, including a cell level energy density of 500 Wh/kg and 1000 cycles with 80% of capacity retention under fast charging conditions (*15*), can barely be achieved by solely using these approaches. Breaking the current bottleneck requires new solutions that can perfect Li deposition on top of these achievements.

In addition to promoting the mass transport, pressurizing the electrode stack has been widely used in modern Li-ion batteries to improve cycling performance by minimizing the interfacial and transport impedance. For Li metal anode, it is qualitatively believed that increasing uniaxial stack pressure helps to alleviate Li dendrite formation and improve CE and cycling performance (*16*). This offers a new possibility to tune the Li morphology beyond the aforementioned strategies promoting mass-transport. Moli Energy mentioned in their patent in 1985 that Li deposits formed under stack pressure showed a denser morphology with enhanced cycling efficiency (*17*). Wilkinson *et al.* (*18*) examined the effect of stack pressure in Li/MoS$_2$ prismatic cells and attributed the Li deformation to the competition between the applied pressure and mechanical strength (creep strength and tensile strength) of the Li. Recent work further proved stack pressure can effectively improve the cycling efficiency and cycle life in anode-free cells (*19–21*), and achieved close-packed morphology (*21*). Undoubtably, applying stack pressure has been extensively proven as an effective method to control the Li deposition morphology. The mechanical properties of Li metal have also been widely studied accordingly (*22–25*). However, the underlying scientific principle of pressure on Li deposition and dissolution behavior at micro/nano scales and how stack pressure can be utilized to control the Li deposition and stripping have not been successfully quantified nor understood. How to achieve the ideal morphology of Li deposits, which consist of large Li particles seamlessly packed on the electrode with very small surface area, and how it can be reversibly deposited and stripped, remain elusive. Answering these questions by establishing a pressure-morphology-performance correlation with optimized Li morphology will open new opportunities to rationally achieve the demanding goals for commercially viable high-energy rechargeable Li metal batteries under various environmental and operating conditions.

Here, combining 3D cryogenic focused ion beam-scanning electron microscopy (3D cryo-FIB-SEM), cryogenic transmission electron microscopy (cryo-TEM), titration gas chromatography (TGC) (*4*), and molecular dynamics (MD) simulation, we elucidated how stack pressure can be applied to precisely manipulate Li deposition and dissolution towards high CE rechargeable Li metal batteries, overcoming the mass-transport bottleneck. Through systematic study of the effects of applied stack pressure on the physical morphology and chemical components of Li deposits, we identified two ways in which the stack pressure regulates the Li nucleation and growth: tuning the favorable Li growth direction at microscale by altering the surface energy at the Li top surface, and densifying Li deposits at nanoscale by exerting mechanical constraints. We found the stack pressure induces negligible impacts on SEI structure and components. In the stripping process, the stack pressure plays a key role in retaining the electronically conductive pathway and minimizing the inactive Li formation, while electrochemically deposited Li reservoir is key to maintaining the dense Li structure and its reversibility upon cycling. Based on the quantitative understanding, we achieved a unprecedent dense Li deposition (99.49% electrode density) with an ideal columnar morphology, minimal surface area, and made it highly reversible upon cycling with minimal inactive Li formation, and thus improved CE (> 99%) at fast charging condition (4 mA/cm$^2$) and room temperature. Such pressure-tailored highly reversible Li metal anodes can be the final push to unlock the potential of high-energy Li metal batteries for fast charging and low temperature operation.

**Pressure effects on Li deposition**

We used a customized split cell with a load cell (Fig. 1A) to precisely control the uniaxial stack pressure applied to the battery during cycling. The pressure was set as the on-set value for the electrochemical performance testing. Figure 1B shows the first cycle CE of Li-Cu cells as a function of applied stack pressure under different current densities from 1, 1.5, to 2 mA/cm$^2$, using ether-based bisalt electrolyte (*26*). At 0 kPa, the CE deceased from 92.5% at 1 mA/cm$^2$ to 85.5% at 2 mA/cm$^2$. When the stack pressure is slightly increased to ~35 kPa, the CE increased for all current densities while the CE at 2 mA/cm$^2$ jumped to 92%. At 350 kPa, the CE was boosted to 98%, 97% and 96% at 1, 1.5 and 2 mA/cm$^2$, respectively. Increasing the stack pressure above 350 kPa cannot further improve the CE. Figure 1C shows the electrochemically deposited Li at a high current density of 2 mA/cm$^2$ for 4 mAh/cm$^2$ exhibits a metallic silver color.

Li-Cu pouch cells were used to test the pressure effects on long-term cycling performances. Figure S1A shows that a nearly doubled cycle life (116 -125 cycles) was achieved for the cells tested under 350 kPa than those (~73 cycles) under 70 KPa, when setting the overpotential limit to – 0.5 V within 30 minutes as the end-of-life condition. In addition, the average CE was improved from ~98% to above 99% by increasing pressure from 70 kPa to 350 kPa, at a high current density of 4 mA/cm$^2$ at room temperature (Fig. S1B). All these results confirm that the optimized stacking pressure plays a critical role in improving the CE and cycling performances of Li metal anode under fast charging conditions.

We then used cryo-FIB-SEM to examine the deposited Li morphology under four representative pressures: 0, 70, 210 and 350 kPa. A high current density of 2 mA/cm$^2$ was applied for the one-hour Li deposition (2 mAh/cm$^2$) morphological study. At 0 kPa, highly porous and whisker-like Li deposits were formed even when using the ether-based electrolyte, as shown in Fig. S2A (top view) and S2B (cross-section). The Li deposits become notably close-packed with increased pressure from 70 kPa to 350 kPa (Fig. 1D-G). The cross-section evolution is even more noticeable. As shown in Fig. 1H-K, along with the increased stack pressure, the electrode thickness obviously decreased. Especially, the cross-section morphology at 350 kPa (Fig. 1K) shows that the Li deposits form perfect columnar structures with large granular diameter of ~4 µm, near-theoretical thickness (9.64 µm, 2 mAh/cm$^2$) of ~10 µm and minimum electrode-level porosity, indicating that stack pressure can be used to precisely control the Li deposition morphology. Further increasing the deposition amount to 4 mAh/cm$^2$, which is required for a practical high-energy battery, the dense, columnar morphology is well maintained (Fig. S3). We predicted in our previous study that the columnar Li deposits is ideal to improve the CE of Li metal by reducing the isolated metallic Li formation (*4*). This study shows that the columnar Li deposits can be achieved by optimizing stack pressure.

It is worth noting that the bottom section of the Li deposits turns from relatively porous at 70 kPa (Fig. 1H) to completely dense at 350 kPa (Fig. 1K), though the top section of the Li deposits at the four different pressures are all dense, indicating the pressure effect plays an important role at the initial stage of Li nucleation. With this assumption, we examined the pressure effects on Li nucleation and initial growth stage with reduced Li deposition loading at 2 mA/cm$^2$ for 0.33 mAh/cm$^2$ under 70, 140, 210 and 350 kPa, respectively. As shown in Fig. 1L-O, the as formed Li nuclei show similar morphology as the bottom part of the one-hour deposits shown in Fig. 1H-K.

We further used cryo-FIB 3D reconstruction to quantify the porosity and volume of Li deposits formed under 70 kPa and 350 kPa (Supplementary Movie S1-2 and Fig.S4). Ideally, the total deposited Li (0.333 mAh/cm$^2$) should exhibit a theoretical thickness of 1.620 µm with zero porosity. When plating at 70 kPa and 350 kPa, the Li layer thickness is measured to be 3.677 µm and 1.697 µm, respectively (Fig. 1P); the porosity is calculated to be 43.57% and 0.51%, respectively (Fig. 1Q). Based on these numbers, the pure deposited Li volumes at 70 kPa and 350 kPa are normalized as 1.107 and 1.036, respectively, which exceed the theoretical value of 1 (Fig. 1R). The increased volume is ascribed to the porous electrode structure, where more Li deposits are exposed to liquid electrolyte and form SEI with large surface areas. Eliminating the porosity of Li deposits is essential to minimize the surface exposure to liquid electrolyte that causes extra parasitic reactions which consume electrolyte and active Li.

Based on the above pressure-tailored Li deposition, we explored the possibility to overcome the mass transport limitations at high rate and low temperature by applying stack pressure: at higher plating rate of 4 mA/cm$^2$ and room temperature, the densely packed columnar structure is still maintained under 350 kPa (Fig. S5); at 0 $^\circ$C, very dense Li deposition can be achieved at 2 mA/cm$^2$ under increased stack pressure of 420 kPa (Fig. S6). These results indicate that applying an optimized stack pressure is a highly feasible way to enable fast charging and low temperature operation for rechargeable Li metal batteries.

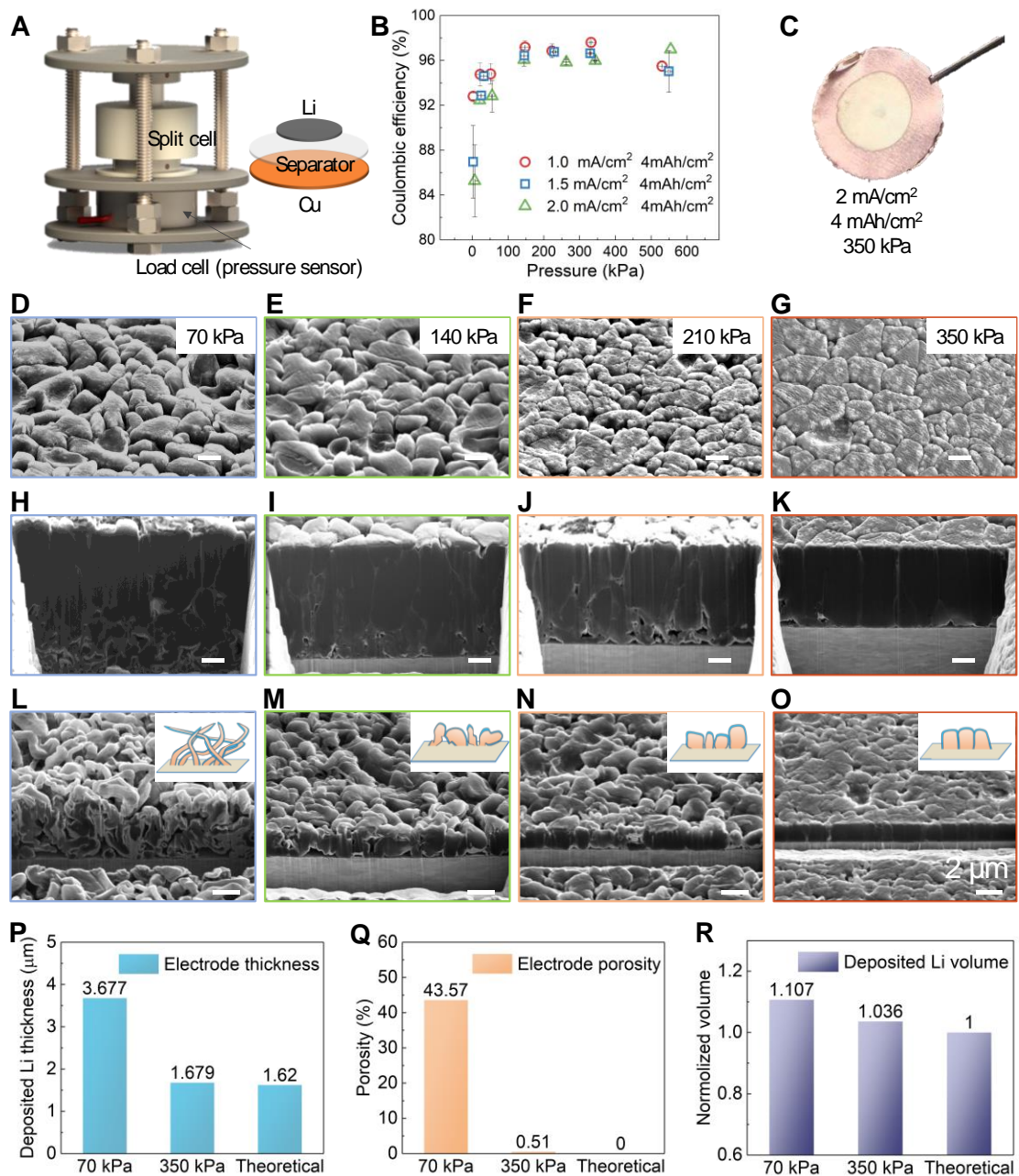

**Fig. 1. Quantifying pressure effects on Li metal anode Coulombic efficiency and plating morphology.** (**A**) the pressure experiment set-up. (**B**) pressure vs. Coulombic efficiency under various current densities. (**C**) Optical image of deposited Li under high current density (2mA/cm$^2$), high loading (4 mAh/cm$^2$), and optimized pressure conditions (350 kPa). (**D-G**) top view and (**H-K**) cross-section of Li deposited under various pressure at 2 mA/cm$^2$ for 1 hour. (**D, H**) 70 kPa, (**E, I**) 140 kPa, (**F, J**) 210 kPa, (**G, K**) 350 kPa. (**L-N**) cross-section SEM images of Li deposits under pressure of (**L**) 70 kPa, (**M**) 140 kPa, (**N**) 210 kPa, (**O**) 350 kPa at 2 mA/cm$^2$ for 10 min (0.333 mAh/cm$^2$). (**P**) electrode thickness, (**Q**) electrode porosity and (**R**) normalized volume of pure deposited Li calculated from 3D cryo-FIB-SEM reconstruction.

MD simulations were applied to reveal the pressure effects on early temporal evolution of Li deposition on Cu surface at nanoscale. We compared the scenarios under 0 kPa (Fig. 2A) and 350 kPa (Fig. 2B). At 0 kPa, the Li deposition began with randomly distributed Li nucleation sites (0.25 ns), evolved as isolated reefs (0.5 ns), grew in an uncontrolled fashion (0.75 ns), and led to a porous morphology with poor surface coverage, uneven thickness and poor interconnectivity (1 ns, see top view evolution in Fig. S7A-D). At 350 kPa, the Li nucleation (0.25 ns) and the promoted connectivity of Li nucleation sites (0.5 ns) created a Li deposition with better homogeneity (0.75 ns) and densified layer (1 ns, see top view evolution in Fig. S7E-H). Better surface area coverage by Li deposits (Fig. S8A) and higher ordering of the Li deposit under stack pressure is also shown by the subtle differences in the short-range Li-Li pairwise distribution function (Fig. S8B). MD simulation reveals stack pressure plays an important role in the temporal evolution of the Li deposition by promoting the lateral Li deposition and densifying the individual Li particle through smoothing the surfaces and eliminating the voids at atomic scales (Fig. 2C and 2D).

Such distinct Li growth behaviors and mechanisms are depicted in Fig. 2 E-F. Without enough uniaxial stack pressure, Li deposit grows freely at the vertical direction, perpendicular to the current collectors (Fig. 2E). The kinetic regime governs the deposited Li stability and morphology, due to the lower diffusion activation barrier at room temperature (*27*) and temporal freedom before reaching the favorable *fcc*-hollow sites on the Cu surface. Such free-growing Li whiskers have been extensively observed in previous *in-situ/operando* studies *(28-30)*, where no stack pressure was present in their experimental set-up. Under the stack pressure, the nucleation and initial growth of the Li deposits adopt a lateral growth along the surface of the current collector (Fig. 2F), due to the free energy change induced by the compressive stress at the electrolyte/separator interface (*31*). He *et al* observed the lateral growth phenomenon using *in-situ* TEM with atomic force microscopy (AFM) applied constraint (*31*). In our case, at the critical pressure when the resistance at the interface exceeds the surface energy of growing laterally, the Li deposits turn to initially grow laterally to fill the intergranular voids, followed by growing at the interface vertically due to the limitation of space laterally and thus form the columnar structure (Fig. 2F). In this way, Li deposits with densely packed columnar morphology can be achieved.

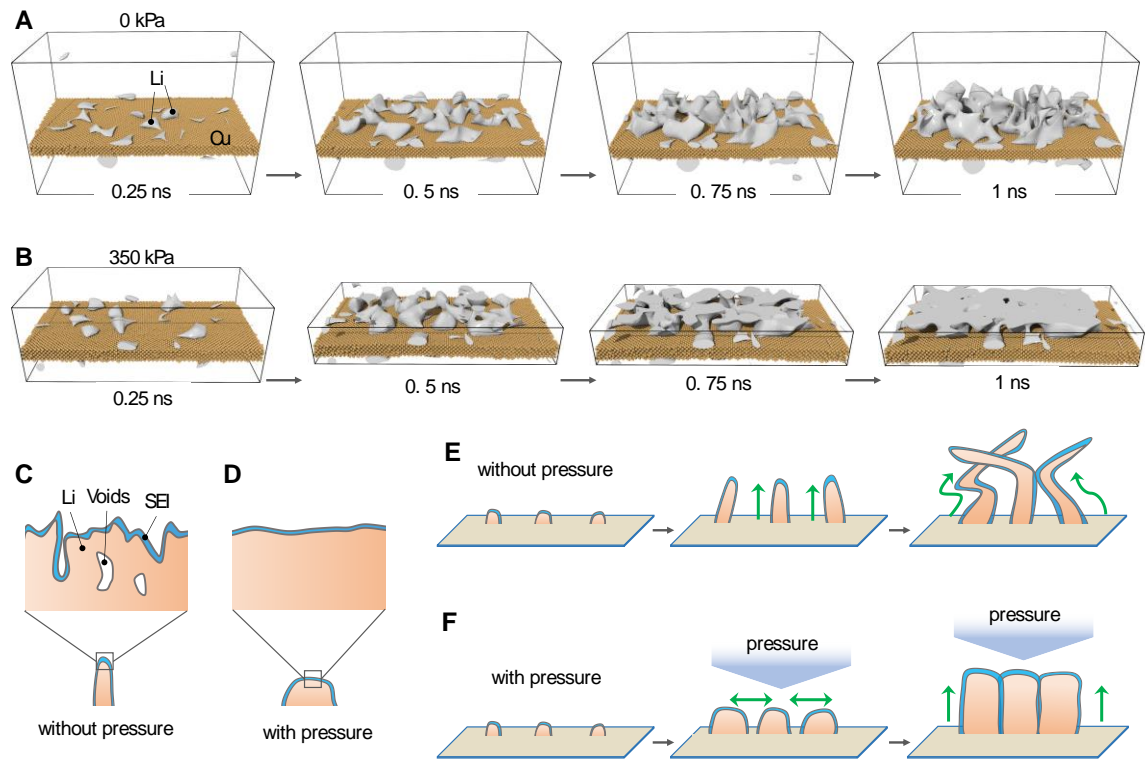

**Fig. 2. MD simulation and schematic illustration of pressure effects on Li nucleation and growth.** The temporal evolution of Li deposition (**A**) at 0 kPa and (**B**) 350 kPa obtained with MD simulations. The cross-section of the Cu surface used for Li deposition is $25.56 \times 12.77$ nm$^2$ with a deposition rate of 20 Li/ps. Additional simulation details can be found in the supplementary material. Atomic-level morphology of Li under (**C**) no stack pressure and (**D**) optimal stack pressure simulated by MD simulation. (**E**) Li nucleation, initial growth and growth under no stack pressure. (**F**) Li nucleation, initial growth and growth under optimal stack pressure. The green arrows indicate the Li growth direction.

## Pressure effects on SEI properties

We then used cryo-TEM to investigate the pressure effects on the SEI structure and components. We comparatively studied the Li formed under 70 kPa and 350 kPa, plating at 2 mA/cm$^2$ for 5 minutes in the ether-based bisalt electrolyte. The Li deposits exhibit a whisker-like morphology at 70 kPa (Fig. 3A) and large granular morphology at 350 kPa (Fig. 3D), in accordance with the micro morphology observed by SEM in Fig. 1L and 1O. Under both stack pressure conditions, we observed the SEI structures and components are almost identical. Figure 3B and 3E compare the nanostructure of the Li deposits under 70 kPa and 350 kPa at large scales. Further zooming in, as shown in Fig. 3C and 3F, the SEI thickness in both samples is 20 - 25 nm, with polycrystalline Li$_2$O embedded on amorphous matrix, showing a Mosaic-type structure. More representative locations for both samples are shown in Fig. S9. The cryo-TEM observation indicates the stack pressure has minimum effects on the SEI structures, components, and their distributions. It primarily affects the Li nucleation and growth processes.

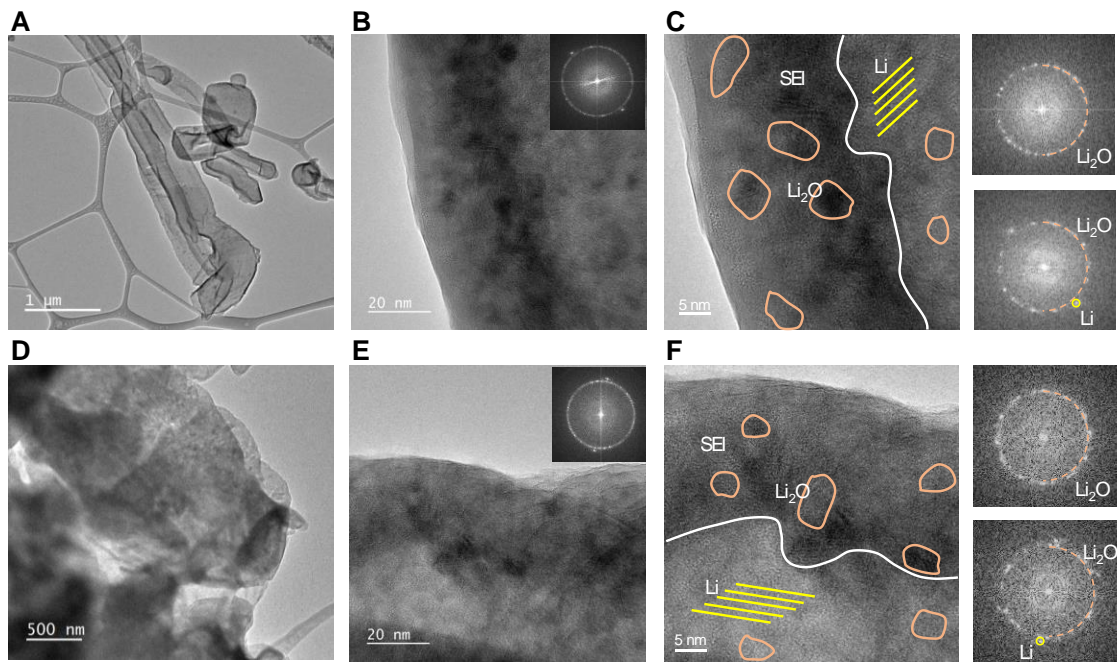

**Fig. 3. Pressure effects on SEI properties by cryo-TEM.** (**A-C**) Li deposited at 70 kPa. (**D-F**) Li deposited at 350 kPa. All Li deposited at 2 mA/cm$^2$ for 5 minutes.

**Pressure effects on Li stripping**

Pressure effects on Li stripping were systematically examined starting from the ideal columnar Li deposits formed at 2 mA/cm$^2$ for 1 hour under 350 kPa (Fig. 4A-B). The stripping rate is 2 mA/cm$^2$. When no pressure is applied during the striping, there are a lot of voids formed in between individual Li columns that causes liquid electrolyte to penetrate through the electrode (Fig. 4C). This facilitates the formation of inactive Li as the Li stripping occurs deep at the base of the columnar structure of the Li deposits. After fully stripping the Li to 1 V under no pressure, a significant amount of porous inactive Li remains on the current collector (Fig. 4D). The CE was only 87% with 12% of the deposited Li remained on the current collector in the form of isolated metallic Li measured by TGC (Fig. 4E), despite having started with fully dense Li deposits. When a stack pressure of 350 kPa was applied during stripping, Li dissolution was constrained to the top surface only (Fig. 4F), thus minimizing the exposed surface area and reducing the inactive Li formation, as the electrolyte cannot penetrate into the roots of the dense Li deposits. After fully stripping to 1 V, only 3% of the total capacity remains as the isolated metallic Li on the current collector surface (Fig. 4G), while the CE is significantly improved to 96% (Fig. 4H). The pressure effect on the stripping process for porous Li deposits also shows the same trend, as shown in Fig. S10. These results reveal that applying stack pressure during the stripping process helps to keep the electrode columnar structure integrity under large ion flux. It is essential to limit the Li stripping taking place only at the top surface to prevent inactive Li formation.

Though optimal pressure was applied, inactive Li formation is still noticeable after fully stripping (Fig. 4I), due to the inevitable inhomogeneity of electrodeposited Li. When fresh Li is further deposited during the following cycle, the columnar structure is hardly maintained (Fig. 4J), ascribing to the interference from the inactive Li residue formed in previous cycles. During extended cycles, more and more inactive Li keeps evolving, breaking the dense morphology (Fig. 4K and 4L), and consuming electrolyte and fresh Li. Significantly, we found if the electrodeposited Li is not fully stripped in each cycle and is partially maintained as a Li reservoir (Fig. 4M), the dense, columnar morphology can be well preserved when Li is re-deposited into the reservoir during extended cycles (Fig. 4N-P). This process is enabled by following the lowest-energy Li diffusion pathway and refilling the existing SEI established during previous cycle. The electrodeposited Li reservoir serves as the re-nucleation sites. In this way, minimum electrolyte and fresh Li will be consumed by continuous inactive Li formation. We further compared the re-plating Li morphology with 1/16, 1/8 and 1/4 of Li reservoir, and identified 1/4 reservoir is essential to maintain the dense morphology (Fig. S11). This study also well explains why a Li-reservoir testing protocol always results in higher CE (*32*), and higher discharge cut-off voltage in a full cell leads to less inactive Li formation (*33*).

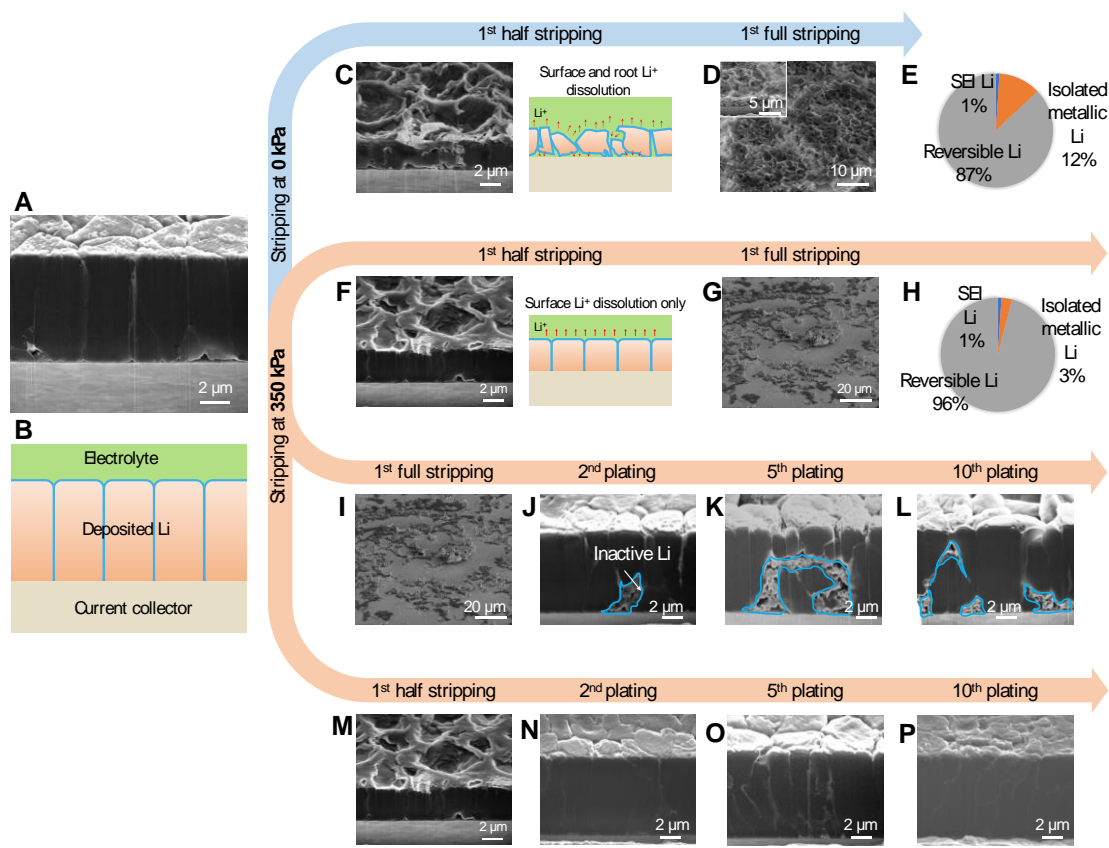

**Fig. 4. Pressure effect on Li stripping process.** (**A, B**) cryo-FIB-SEM image and schematic illustration of columnar Li plated at 350 kPa. (**C-E**) Li stripping at 0 kPa: (**C**) cross-section morphology of half-stripped Li; (**D**) fully stripped Li; (**E**) capacity usage analysis by TGC. (**F-H**) Li stripping at 350 kPa: (**F**) cross-section morphology of half-stripped Li; (**G**) fully stripped Li;

(**H**) capacity usage analysis by TGC. (**I-P**) **Li reservoir effect study**, plating and stripping under stack pressure of 350 kPa: (**I-L**) Li deposition morphology evolution using full-stripping protocol for 10 cycles. (**M-P**) Li deposition morphology evolution using half-stripping protocol to retain Li reservoir for 10 cycles. All plating and stripping at 2 mA/cm$^2$, plating for 1 hour, half stripping for 30 min, full stripping to 1V.

In summary, we identified that the stack pressure can be used a powerful tuning knob to precisely tailor Li deposition morphology and dissolution geometry. Using multiscale characterization tools, we discovered that applying optimized stack pressure can fine tune Li nucleation and growth direction towards dense deposition, staying away from the dendrite growth caused by mass transport limitations. We achieved the predicted ideal columnar Li deposit with minimal electrode porosity by optimizing the on-set stack pressure at 350 kPa. During the Li stripping process, pressure assures the close interfacing between the dense Li deposits and current collector to prevent the liquid electrolyte from penetrating into the root of the columnar structure, thus dramatically reducing the inactive Li formation. The electrochemically formed dense Li reservoir is the key to maintain the columnar structure reversible upon extended cycling, greatly improving the cycle life. Such unprecedented manipulation of battery electrochemical behavior using stack pressure represents a critical step towards new design rules and new manufacturing process enable practical Li metal batteries and other metal anodes.

**References and notes**

1. X. B. Cheng, R. Zhang, C. Z. Zhao, Q. Zhang, Toward safe lithium metal anode in rechargeable batteries: a review. *Chem. Rev.* **117**, 10403–10473 (2017).

2. C. Fang, X. Wang, Y. S. Meng, Key Issues Hindering a Practical Lithium-Metal Anode. *Trends Chem.* **1**, 152–158 (2019).

3. M. Winter, B. Barnett, K. Xu, Before Li Ion Batteries. *Chem. Rev.* **118** (2018), pp. 11433–11456.

4. C. Fang *et al.*, Quantifying inactive lithium in lithium metal batteries. *Nature*. **572**, 511–515 (2019).

5. W. Xu *et al.*, Lithium metal anodes for rechargeable batteries. *Energy Environ. Sci.* **7** (2014), pp. 513–537.

6. J. N. Chazalviel, Electrochemical aspects of the generation of ramified metallic electrodeposits. *Phys. Rev. A*. **42**, 7355–7367 (1990).

7. J. Xiao, How lithium dendrites form in liquid batteries. *Science* **366**, 426–427 (2019).

8. X. Cao *et al.*, Monolithic solid–electrolyte interphases formed in fluorinated orthoformate-based electrolytes minimize Li depletion and pulverization. *Nat. Energy*. **4**, 796–805 (2019).

9. Y. Yang *et al.*, High-Efficiency Lithium-Metal Anode Enabled by Liquefied Gas

Electrolytes. *Joule*. **3**, 1986–2000 (2019).

10. S. Chen *et al.*, High-Voltage Lithium-Metal Batteries Enabled by Localized High-Concentration Electrolytes. *Adv. Mater.* **1706102**, 1–7 (2018).

11. C. Niu *et al.*, Self-smoothing anode for achieving high-energy lithium metal batteries under realistic conditions. *Nat. Nanotechnol.* **14**, 594–601 (2019).

12. D. Cao *et al.*, 3D Printed High-Performance Lithium Metal Microbatteries Enabled by Nanocellulose. *Adv. Mater.* **31**, 68–71 (2019).

13. R. Xu *et al.*, Artificial Interphases for Highly Stable Lithium Metal Anode. *Matter*. **1**, 317–344 (2019).

14. J. Wang *et al.*, Improving cyclability of Li metal batteries at elevated temperatures and its origin revealed by cryo-electron microscopy. *Nat. Energy*. **4**, 664–670 (2019).

15. J. Liu *et al.*, Pathways for practical high-energy long-cycling lithium metal batteries. *Nat. Energy*. **4**, 180–186 (2019).

16. T. Hirai, Influence of Electrolyte on Lithium Cycling Efficiency with Pressurized Electrode Stack. *J. Electrochem. Soc.* **141**, 611 (1994).

17. K. Brandt, J. A. R. Stiles, Battery and methods of making the battery (1985).

18. D. P. Wilkinson, H. Blom, K. Brandt, D. Wainwright, Effects of physical constraints on Li cyclability. *J. Power Sources*. **36**, 517–527 (1991).

19. X. Yin *et al.*, Insights into morphological evolution and cycling behaviour of lithium metal anode under mechanical pressure. *Nano Energy*. **50**, 659–664 (2018).

20. A. J. Louli *et al.*, Exploring the Impact of Mechanical Pressure on the Performance of Anode-Free Lithium Metal Cells. *J. Electrochem. Soc.* **166**, 1291–1299 (2019).

21. R. Weber *et al.*, Long cycle life and dendrite-free lithium morphology in anode-free lithium pouch cells enabled by a dual-salt liquid electrolyte. *Nat. Energy* (2019), doi:10.1038/s41560-019-0428-9.

22. X. Zhang *et al.*, Rethinking How External Pressure Can Suppress Dendrites in Lithium Metal Batteries. *J. Electrochem. Soc.* **166**, 3639–3652 (2019).

23. A. Masias, N. Felten, R. Garcia-Mendez, J. Wolfenstine, J. Sakamoto, Elastic, plastic, and creep mechanical properties of lithium metal. *J. Mater. Sci.* **54**, 2585–2600 (2019).

24. C. Xu, Z. Ahmad, A. Aryanfar, V. Viswanathan, J. R. Greer, Enhanced strength and temperature dependence of mechanical properties of Li at small scales and its implications for Li metal anodes. *Proc. Natl. Acad. Sci. U. S. A.* **114**, 57–61 (2017).

25. Y. Wang, D. Dang, X. Xiao, Y. T. Cheng, Structure and mechanical properties of electroplated mossy lithium: Effects of current density and electrolyte. *Energy Storage Mater*. **26**, 276–282 (2020).

26. J. Alvarado *et al.*, Bisalt ether electrolytes: A pathway towards lithium metal batteries with Ni-rich cathodes. *Energy Environ. Sci.* **12**, 780–794 (2019).

27. D. Gaissmaier, D. Fantauzzi, T. Jacob, First principles studies of self-diffusion processes on metallic lithium surfaces. *J. Chem. Phys*. **150**, 41723 (2019).

28. P. Bai, J. Li, F. R. Brushett, M. Z. Bazant, Transition of lithium growth mechanisms in liquid electrolytes. *Energy Environ. Sci. Energy Environ. Sci*. **9**, 3221–3229 (2016).

29. H. Ghassemi, M. Au, N. Chen, P. A. Heiden, R. S. Yassar, Real-time observation of lithium fibers growth inside a nanoscale lithium-ion battery. *Appl. Phys. Lett*. **99**, 123113 (2011).

30. Z. Zeng *et al.*, Visualization of electrode-electrolyte interfaces in LiPF6/EC/DEC electrolyte for lithium ion batteries via in situ TEM. *Nano Lett*. **14**, 1745–1750 (2014).

31. Y. He *et al.*, Origin of lithium whisker formation and growth under stress. *Nat. Nanotechnol*. **14**, 1042–1047 (2019).

32. B. D. Adams, J. Zheng, X. Ren, W. Xu, J. G. Zhang, Accurate Determination of Coulombic Efficiency for Lithium Metal Anodes and Lithium Metal Batteries. *Adv. Energy Mater*. **1702097**, 1–11 (2017).

33. A. J. Louli *et al.*, Diagnosing and correcting anode-free cell failure via electrolyte and morphological analysis. *Nat. Energy* (2020), doi:10.1038/s41560-020-0668-8.

**Acknowledgments:** We thank Dr. Jinxing Li for his valuable suggestions on the manuscript. **Funding:** This work was supported by the Office of Vehicle Technologies of the U.S. Department of Energy through the Advanced Battery Materials Research (BMR) Program (Battery500 Consortium) under Contract DE-EE0007764. Cryo-FIB was performed at the San Diego Nanotechnology Infrastructure (SDNI), a member of the National Nanotechnology Coordinated Infrastructure, which is supported by the National Science Foundation (grant ECCS-1542148). We acknowledge the UC Irvine Materials Research Institute (IMRI) for the use of the cryo-TEM, funded in part by the National Science Foundation Major Research Instrumentation Program under Grant CHE-1338173; **Author contributions:** C.F. and Y.S.M. conceived the ideas. C.F. designed the experiments. Bi.L. implemented the electrochemical tests. Bi.L., C.F. and D.C. performed the cryo-FIB experiments. G.P. and Bo.L. performed the MD simulations. M.Z. collected the cryo-TEM data. C.F. conducted TEM data interpretation. S.C. and M.C. conducted the pouch cell tests. M.C. prepared electrolytes. J.M.D. conducted the load cell design and calibration. C.F. wrote the manuscript. All authors discussed the results and commented on the manuscript. All authors have given approval to the final version of the manuscript; **Competing interests:** Authors declare no competing interests; **Data and materials availability:** All data is available in the main text or the supplementary materials.